\documentclass[twocolumn,pra,showpacs,nofootinbib,superscriptaddress]{revtex4}
\usepackage{amssymb}
\usepackage{dcolumn,bm,graphicx,amsmath}

\begin{document}

\preprint{UNR Jan 2005-\today }
\title{Multipolar theory of black-body radiation shift of atomic energy
levels and its implications for optical lattice clocks}
\author{Sergey G. Porsev}
\affiliation{Department of Physics, University of Nevada, Reno, Nevada 89557}
\affiliation{Petersburg Nuclear Physics Institute, Gatchina 188300, Russia}
\author{Andrei Derevianko}
\affiliation{Department of Physics, University of Nevada, Reno, Nevada 89557}
\date{\today}

\begin{abstract}
A black-body radiation (BBR) shifts of $nsnp\,^3\!P_0 - ns^2\,^1\!S_0$ clock
transition in divalent atoms Mg, Ca, Sr, and Yb are evaluated. A theory of
multipolar BBR shifts is developed and its implications are discussed. At
room temperatures, the resulting uncertainties in the BBR shifts are
relatively large and substantially affect the projected $10^{-18}$
fractional accuracy of the optical-lattice-based clocks.
\end{abstract}

\maketitle

Atomic clocks based on ultranarrow $^3\!P_0 - ^1\!S_0$ transition in
divalent atoms may offer a new level of time-keeping accuracy. In
this scheme the atoms are confined in optical lattice. The lattice
laser wavelength is selected in such a way that the dominant
perturbation of the clock frequency, the induced ac Stark shifts,
for both clock states exactly
cancel. Although other effects still perturb the clock frequency, estimates~%
\cite{TakHinHig05} indicate that the projected fractional
uncertainty of such clocks may be as low as $10^{-18}$. By
comparison, $10^{-15}$ is the fractional uncertainty of the current
Cs standard realizing the SI definition of the unit of time. This
apparent advantage of the optical-lattice clocks has motivated a
number of recent proposals: original Katori's scheme~\cite{Kat02}
with fermionic Sr isotopes has been extended to
Mg~\cite{ThomsenPrivate}, Ca~\cite{RieBinWil03}, and
Yb~\cite{PorDerFor04} atoms and to bosonic
isotopes~\cite{SanAriIdo05,HonCraNag05}. In addition, various
schemes of probing the highly-forbidden $nsnp\,^3\!P_0
-ns^2\,^1\!S_0 $ clock transition have been proposed: three-photon
transition, electromagnetically-induced transparency, and transition
assisted by external magnetic
field~\cite{SanAriIdo05,HonCraNag05,TaiYudOat05}.

Considering advantages of optical lattice clocks, here we investigate an
important systematic effect of the black-body radiation (BBR) on the
frequency of the $^3\!P_0-^1\!S_0$ clock transition. Indeed, the SI
definition of the second explicitly involves atomic clock operating at the
absolute zero of temperature. In a laboratory environment with an ambient
temperature $T$, one needs to introduce the $T$-dependent BBR correction to
the observed frequency. Here, using techniques of many-body relativistic
atomic structure, we compute the BBR shift for Mg, Ca, Sr, and Yb and
evaluate uncertainties of the calculations. As summarized in Table~\ref%
{Tab:BBRsum}, the resulting fractional uncertainties in the clock
frequencies at $T=300\,\mathrm{K}$ are \emph{large}, ranging from $1
\times 10^{-17}$ for Mg to $3 \times 10^{-16}$ for Yb.

The main conclusions of this paper are (i) the present uncertainty in our
computed BBR shift is an obstacle on the way towards the projected $10^{-18}$
accuracy goal; (ii) due to $T^4$ scaling of the BBR shift, it may be
beneficial to operate at low temperatures, e.g., at liquid nitrogen
temperatures; (iii) if operating at room temperatures, high-precision
(0.02\%-accurate for Sr) measurements of the BBR shifts or related
quantities are required; (iv) Mg-based clock is the least susceptible to
BBR; compared to Sr, the Mg BBR shift is an order of magnitude smaller (see
Table~\ref{Tab:BBRsum}). Additionally, we develop a general relativistic
theory of the BBR shift caused by multipolar (EJ and MJ) components of the
radiation field.

\begin{table}[h]
\caption{Black-body radiation shift for clock transitions between the
lowest-energy $^3\!P_0$ and $^1\!S_0$ states in divalent atoms. $\protect%
\delta\protect\nu_{\mathrm{BBR}}$ is the BBR shift at $T=300\,\mathrm{K}$
with our estimated uncertainties. $\protect\nu_0$ is the clock transition
frequency, and $\protect\delta\protect\nu_{\mathrm{BBR}}/\protect\nu_0$ is
the fractional contribution of the BBR shift. The last column lists
fractional errors in the absolute transition frequencies induced by the
uncertainties in the BBR shift. }
\label{Tab:BBRsum}%
\begin{tabular}{cdccc}
\hline \hline
Atom
&\multicolumn{1}{r}{$\delta\nu_{\mathrm{BBR}}$, Hz}
&\multicolumn{1}{c}{$\nu_0$, Hz}
&\multicolumn{1}{c}{$\delta\nu_{\mathrm{BBR}}/\nu_0$}
&\multicolumn{1}{c}{uncertainty}\\
\hline
Mg   &  -0.258(7)  & $6.55 \times 10^{14}$ & $-3.9  \times 10^{-16}$ & $1  \times 10^{-17}$\\
Ca   &  -1.171(17) & $4.54 \times 10^{14}$ & $-2.6  \times 10^{-15}$ & $4  \times 10^{-17}$\\
Sr   &  -2.354(32) & $4.29 \times 10^{14}$ & $-5.5  \times 10^{-15}$ & $7  \times 10^{-17}$\\
Yb   &  -1.25(13)  & $5.18 \times 10^{14}$ & $-2.4  \times 10^{-15}$ & $3  \times 10^{-16}$ \\
\hline \hline
\end{tabular}
\end{table}

The paper is organized as follows. Firstly, we derive relativistic formulae
for the BBR shift which incorporate various multipolar contributions of the
radiation field. Secondly, we apply these general expressions to
determination of the BBR shifts in optical-lattice clocks with Mg, Ca, Sr,
and Yb atoms. Unless specified otherwise, atomic units and the Gaussian
system of electromagnetic units are used throughout. In these units, $%
c=1/\alpha$, where $\alpha\approx 1/137$ is the fine-structure constant.
Temperature is expressed in units of $E_h/k_B$, where $E_h$ is the Hartree
energy and $k_B$ is the Boltzmann constant.

\emph{Multipolar theory of the black-body radiation shift.} The BBR shift is
caused by perturbation of the atomic energy levels by the oscillating
thermal radiation. Both atomic levels involved in the clock transition are
perturbed and the overall BBR correction is a difference of the BBR shifts
for the two levels. We find that determining shift for the upper $^3\!P_0$
level requires certain care. This level is a part of the $^3\!P_J$
fine-structure manifold, $J=0,1,2$. The separation between the levels in the
manifold is comparable to the characteristic wavenumber of the BBR
radiation, $208.51 \, \mathrm{cm}^{-1}$, at $T=300\,\mathrm{K}$, and
contributions of the BBR-induced magnetic-dipole and electric-quadrupole
transitions to the levels of the manifold may be enhanced. Taking these
induced transitions into account requires going beyond the conventional
electric-dipole approximation, Ref.~\cite{FarWin81}.

Considering a potential importance of the multipolar contributions, in this
section we derive the relevant formulae for BBR-induced energy shifts.
Although we develop a general relativistic theory including retardation, at
the end of the section we reduce our expressions to practically important
non-relativistic non-retarded formulas. The derived expressions are
generalization of the electric-dipole BBR shift of Ref.~\cite{FarWin81}.

The BBR spectral density is given by the Plank formula
\begin{equation*}
u_\omega\left( T \right) =\frac{\alpha^3}{\pi^{2}}\omega^{3}\frac{1}{%
\exp\left( \omega/T\right) -1}.
\end{equation*}
It is a weak perturbation and the time-evolution of the reference state $%
|g\rangle$ for off-resonance excitations can be computed assuming that the
excited state amplitudes adiabatically follow that of the reference state.
With a generalization to narrow resonant contributions~\cite{FarWin81}, the
BBR shift is given by
\begin{eqnarray}
\delta E_{g}&=&\frac{1}{4\alpha^2}\sum_{\varepsilon}\int d\hat{k} \, \mathrm{%
P.V.}\!\int_0^\infty u_\omega(T)\,\frac{d\omega}{\omega^2}
\label{Eq:EgBBRGeneral} \\
&&\times \sum_{p}\left\{ \frac{\left[ h^{\left( +\right) }\right] _{gp}\left[
h^{\left( -\right) }\right] _{pg}}{\omega_{gp}+\omega}+\frac{\left[
h^{\left( -\right) }\right] _{gp}\left[ h^{\left( +\right) }\right] _{pg}}{%
\omega _{gp}-\omega}\right\} \, ,  \notag
\end{eqnarray}
with averaging over photon propagation directions $\hat{k}$ and
polarizations $\bm{\varepsilon}$. The second-order summation is over
intermediate atomic states $|p\rangle$ and involves the Coulomb-gauge
couplings $h^{\left( \pm\right) }=\left( \bm{\alpha}\cdot\bm{\varepsilon}%
\right) \exp\left( \mp i\left( \mathbf{k}\cdot\mathbf{r}\right) \right)$, $%
\bm{\alpha}$ encapsulating the conventional Dirac matrices. P.V. denotes the
Cauchy's principal value; as elucidated in Ref.~\cite{FarWin81} it is
required for a proper treatment of resonant contributions.

While evaluating matrix elements of operators $h^{\left( \pm \right) }$, we
use multipolar expansion of $e^{i\left( \mathbf{k}\cdot
\mathbf{r}\right) }$ in vector spherical harmonics and express the resulting
couplings in terms of traditional multipole moments $%
Q_{JM}^{\left( \lambda \right) }$

\begin{eqnarray*}
(\bm{\alpha} \bm{\varepsilon}) e^{i\left( \mathbf{k}\cdot \mathbf{r}\right) }
&=&-\sum_{JM\lambda }i^{J+1+\lambda }\left( \mathbf{Y}_{JM}^{\left( \lambda
\right) }\left( \hat{k}\right) \cdot \bm{\varepsilon}\right)  \\
&&\times \sqrt{\frac{4\pi \left( 2J+1\right) \left( J+1\right) }{J}}\frac{%
k^{J}}{\left( 2J+1\right) !!}Q_{JM}^{\left( \lambda \right) }.
\end{eqnarray*}%
Here $\lambda =0$ is for magnetic (MJ) and $\lambda =1$ is for
electric (EJ) multipolar amplitudes. Explicit relativistic
expressions with retardation for matrix elements of $Q_{JM}^{\left(
\lambda \right) }$ can be found in Ref.~\cite{JohPlaSap95}.
Neglecting retardation effects (i.e., in the long-wavelength approximation) $%
Q_{JM}^{\left( 1\right) }$ become frequency-independent EJ moments $%
Q_{JM}^{\left( 1\right) }=r^{J}C_{JM}\left( \hat{r}\right)$, where
$C_{JM}\left( \hat{r}\right)$ are normalized spherical harmonics. In
the case of magnetic-dipole transitions in the nonrelativistic limit
$Q_{1M}^{\left( 0\right) }=-\frac{\alpha }{2}\left(
\mathbf{L}+2\mathbf{S}\right) _{M}$. Notice that the retardation
brings correction in the order of $(\alpha \omega )^{2}$ to these
expressions.

Averaging over polarizations and propagation directions in Eq.~(\ref%
{Eq:EgBBRGeneral}), we find that the BBR shift is a sum over 
multipolar contributions: $\delta E_{g}=\sum_{J\lambda}\delta E_{g}^{\left(
J\lambda\right) }$,
\begin{eqnarray}
\delta E_{g}^{\left( J\lambda\right) }&=&-\pi\frac{J+1}{J [\left(
2J-1\right)!!]^2} \, \alpha^{2\left( J-1\right) } \\
&& \times \mathrm{P.V.}\! \int_{0}^{\infty }d\omega~\omega^{2\left(
J-1\right) }~u_\omega(T) ~\alpha _{g}^{\left( J\lambda\right) }\left(
\omega\right) \,.  \notag
\end{eqnarray}
Here $\alpha_{g}^{\left( J\lambda\right) }\left( \omega\right) $ are the
generalized dynamic multipolar scalar polarizabilities
\begin{equation}
\alpha_{g}^{\left( J\lambda\right) }\left( \omega\right) = \frac{2}{2J+1}%
\sum_{p,M} |\left\langle p\right\vert Q_{JM}^{\left( \lambda\right) }
\left\vert g\right\rangle|^2 \left\{ \frac{\omega_{pg}}{\omega_{pg}^2-%
\omega^2}\right\} \,.  \label{Eq:polarizGen}
\end{equation}

A cursory examination of these formulas reveals that compared to $2^{J}$
multipole, the contribution of $2^{J+1}$ multipole is suppressed by a factor
of $\alpha^{2}\approx(1/137)^{2}$. Also for the same $J$ the magnetic
contribution is $\alpha^{2}$ weaker than that of the EJ photons. As in the
theory of multipolar radiative transitions E(J+1) and MJ contributions are
of the same order in $\alpha$.

\begin{figure}[h]
\begin{center}
\includegraphics[scale=0.5]{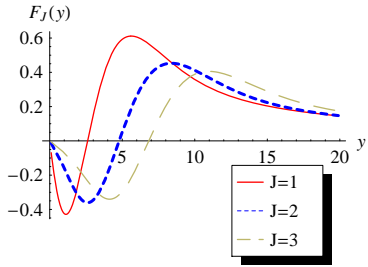}
\end{center}
\caption{Universal multipolar functions $F_{J}(y)$, Eq.~(\protect\ref{Eq:FJ}%
), for $J=1,2,3$. }
\label{Fig:FJ}
\end{figure}

To illuminate the $T$-dependence of contributions of individual
intermediate states we recast the BBR shifts into a form ($J_g$ is
the total angular momentum of the reference state, $\langle
g||Q_{JM}^{\left(\lambda\right)}||p\rangle$ is the reduced matrix
element)
\begin{eqnarray}
\delta E_{g}^{\left( \lambda J\right) }=-\frac{\left( \alpha T\right) ^{2J+1}%
}{2J_{g}+1} \sum_{p} \left\vert \langle g||Q_J^{\left( \lambda\right)
}||p\rangle\right\vert ^{2}F_{J} \left(\frac{\omega_{pg}}{T} \right),
\label{Eq:dEBBRxFJ}
\end{eqnarray}
with universal functions
\begin{eqnarray}
F_{J}\left( y\right) &=&\frac{1}{\pi}\frac{J+1}{J\left( 2J+1\right) !!\left(
2J-1\right) !!}  \label{Eq:FJ} \\
&& \times P.V.\int_{0}^{\infty}\left( \frac{1}{y+x}+\frac {1}{y-x}\right)
\frac{x^{2J+1}}{e^{x}-1}dx \, .  \notag
\end{eqnarray}
Functions $F_{J}\left(y\right)$ are multipolar generalizations of function $%
F\left( y\right) $ introduced by \citet{FarWin81} in the E1 case. We plot
our computed $F_{J}$ functions for $J=1,2,3$ in Fig.~\ref{Fig:FJ}. $F_J(y)$
are odd functions with respect to $y$. From examining Fig.~\ref{Fig:FJ}, it
is clear that $F_J$ rapidly change around $y \sim 1$ and slowly fall off for
$y \gg 1$. Depending on the value of excitation energy, $\omega_{pg} = y \, T
$, a particular intermediate state may introduce either negative or positive
BBR shift. Notice that $F_J$ are broad distributions, they have comparable
values for $|y|\lesssim 20$; this will have implications for interpreting
our results.

At large values of the argument $\left\vert y\right\vert \gg1$, $F_{J}(y)
\propto 1/y$. The limit $y\gg 1$ corresponds to the case when the transition
energy is much larger than $T$. If all virtual transitions satisfy this
requirement, then the leading contribution to the multipolar BBR shift can
be expressed in terms of static polarizabilities
\begin{eqnarray}
\delta E_{g}^{\left( J\lambda\right) } = - \frac{\zeta(2J+2)(2J+2)!}{2\pi J
\, [\left( 2J-1\right)!!]^2} \, \alpha^{2J+1} T^{2J+2} \alpha_g^{(J
\lambda)}(0),  \label{Eq:deltaEg}
\end{eqnarray}
where $\zeta$ is the Riemann zeta-function. As the scaling factor, $%
\alpha^{2J+1} T^{2J+2}$ , is expressed in atomic units, we observe that as
multipolarity~$J$ increases by one, in addition to the usual $\alpha^{2}$
suppression, there is a temperature suppression factor of $\left(
k_{B}T/E_{h}\right)^{2}$. For $T=300 \, \mathrm{K}$ this suppression is
sizable, as $\left( k_{B} T/E_{h}\right)^{2}\approx 9.0\times 10^{-7}$.

\emph{BBR shift for $^3\!P_0-\,^1\!S_0$ transition in divalent atoms}. Below
we apply the developed formalism to computing the BBR shift for the $^1\!S_0
- ^3\!P_0$ clock transition in divalent atoms. We will assume that the atoms
are at the ambient temperature of $T=300\,\mathrm{K}$. Both clock levels
experience the BBR shift and the total shift $\delta\nu_\mathrm{BBR}$ is the
difference between the two individual shifts, $\delta\nu_\mathrm{BBR} =
\delta\nu_\mathrm{BBR}(^3\!P_0) - \delta\nu_\mathrm{BBR}(^1\!S_0)$.

Consider first the BBR shift of the ground $^1\!S_0$ state. Here transition
energies of various multipolar transitions to the upper levels are much
larger than $T$, i.e., we are in the $y\gg1$ limit of Fig.~\ref{Fig:FJ}.
Here compared to the dominant E1-induced shift, the contribution of M1
transitions is suppressed by $\alpha^2 \sim 10^{-4} $ and E2 by $\alpha^2
\left( k_{B} T/E_{h}\right)^{2}\sim 10^{-10}$. Higher-order multipoles are
suppressed even more. As to the retardation effects in E1 matrix elements,
we expect that they would be suppressed by a factor of $\alpha^2 \left(
k_{B} T/E_{h}\right)^{2}\sim 10^{-10}$. Nevertheless, since the fractional
contribution of the BBR shift to the clock frequency is at $5 \times 10^{-15}
$ level (see Table~\ref{Tab:BBRsum}), one would need to introduce the M1
corrections at the projected accuracy of $10^{-18}$.

For the $^3\!P_0$ levels, the characteristic thermal photon frequency is
comparable to the fine-structure intervals for the $^3\!P_J$ manifold. The $%
^3\!P_0$ level is connected by M1 transition to the $^3\!P_1$ level and by
E2 transition to the $^3\!P_2$ level. For these transitions the values of
the relevant functions $F_J\sim 1$, see Fig.~\ref{Fig:FJ}, and we estimate $%
\delta E_{g}^{\left( M1\right)} \sim \alpha^2 \, (\alpha T)^3$,
$\delta E_{g}^{\left( E2\right)} \sim (\alpha T)^5$, while $\delta
E_{g}^{\left( E1\right)} \sim \alpha^3 ( T)^4
/\omega_{^3\!D_1-^3\!P_0}$. Our numerical estimate, based on the
transitions inside the fine-structure manifold lead to the following
values of the BBR shifts for Sr: $\delta E_{g}^{\left( M1\right) } =
2.4 \times 10^{-5}~\mathrm{Hz}$ and $\delta E_{g}^{\left( E2\right)
} = 2.5 \times 10^{-8}~\mathrm{Hz}$. Since the E1 BBR shift for Sr
is $\sim 2\, \mathrm{Hz}$, the M1 and E2 contributions can be
neglected at the present 1\%-level of accuracy of our calculations.

We find that although the thermal photon energy is close to the
fine-structure intervals, the induced multipole BBR shifts are not
amplified. The main reason is that the BBR energy distribution is broad: the
universal functions $F_J$ have comparable values for a wide range of
excitation energies, $|\omega| \lesssim 20 \, T$, see Fig.~\ref{Fig:FJ}. For
example, for Sr $^{3}P_{0}-~^{3}D_{1}~$\ E1 transition $F_{1}\approx0.16$,
while for the $^{3}P_{0}-~^{3}P_{1}$ M1 transition $F_{1}\approx-0.41$ and
for the $^{3}P_{0}-~^{3}P_{2}$ E2 transition $F_{2}\approx-0.36$. For such a
broad distribution, the multipolar BBR shift is determined by the prefactor
in Eq.~(\ref{Eq:dEBBRxFJ}) resulting in a suppression of multipoles beyond
E1.

Based on the above discussion, we may exclusively focus on the
electric-dipole ($J=1,\lambda=1$) contribution to the BBR shift. >From our
general expressions we obtain an approximate formula,
\begin{eqnarray}
&&\delta E_{g}^{(E1)} \approx -\frac{2}{15} (\alpha \pi)^3 T^4
\alpha_g^{(E1)}(0) \times \left[ 1 + \eta \right] \, ,  \label{Eq:delEg} \\
&&\eta = \frac{(80/63) \pi^2}{\alpha_g^{(E1)}(0) T} \sum_p \frac{ |\langle
p||Q_{1}^{\left( 1\right)}||g \rangle|^2 }{(2J_g+1) y_p^3}\left( 1+ \frac{21
\pi^2}{5y_p^2} + \frac{336 \pi^4}{11y_p^4} \right) \, .  \notag
\end{eqnarray}
Here $y_p = \omega_{pg}/T$ and $\alpha_g^{(E1)}(0)$ is the traditional
static dipole polarizability. To arrive at the above equation, we used
asymptotic expansion $F_{1}\left( y\right) \approx \frac{4\pi ^{3}}{45y}+%
\frac{32\pi ^{5}}{189y^{3}}+ \frac{32\pi ^{7}}{45y^{5}}+\frac{512\pi ^{9}}{%
99y^{7}}$, which has an accuracy better than 0.1\% for $|y|>10$. $\eta$
represents a "dynamic" fractional correction to the total shift. The leading
contribution is determined by polarizability and below we compute $%
\alpha_{g}^{(E1)}(0)$ using methods of atomic structure.

Evaluation of the static dipole polarizabilities follows a procedure of Ref.~%
\cite{PorDer06Z}. Here we only outline the major steps. We employ
relativistic many-body code described in Refs.~\cite%
{DzuFlaKoz96b,DzuKozPor98,KozPor99E}. The employed formalism is a
combination of configuration-interaction method in the valence space
with many-body perturbation theory for core-polarization effects. In
this method, one determines wave functions from solving the
effective many-body Schr\"{o}dinger equation
\begin{equation}
\left\{ H_\mathrm{FC} + \Sigma(E) \right\} \, | \Psi_n \rangle = E_n
\, |\Psi_n \rangle
 \label{Heff} \, ,
\end{equation}
Here $H_\mathrm{FC}$ is the frozen-core Dirac-Hartree-Fock
Hamiltonian and self-energy operator $\Sigma$ is a core-polarization
correction. To improve upon this approximation, one can introduce an
adjustable energy shift $\delta $ and replace $\Sigma(E) \rightarrow
\Sigma(E-\delta)$ in the effective Hamiltonian, Eq.~(\ref{Heff}). We
have determined $\delta$ empirically, from a fit of theoretical
energy levels to experimental spectrum. Inclusion of this shift
mimics high-order corrections in perturbation theory. In addition,
we incorporated dressing of the external electromagnetic field (core
shielding) in the framework of the random-phase approximation (RPA).
To find valence contribution to $\alpha_{g}^{(E1)}(0)$ we summed
over the intermediate states in Eq.~(\ref{Eq:polarizGen}) using the
Dalgarno-Lewis-Sternheimer method~\cite{DalLew55}. A small
correction to polarizability due to core-excited intermediate states in Eq.~(\ref%
{Eq:polarizGen}) was computed within the relativistic RPA.

\begin{table}[tbp]
\caption{Static electric dipole polarizabilities in a.u. and BBR shifts in
Hz for the ground $^1\!S_0$ and the lowest-energy $^3\!P_0$ excited states
of Mg, Ca, Sr, and Yb atoms. Theoretical uncertainties are indicated in
parenthesis.}
\label{Tab:polariz}%
\begin{ruledtabular}
\begin{tabular}{ccccc}
                                      &    Mg    &    Ca      &    Sr    &    Yb      \\
\hline
$\alpha_{^1\!S_0}$                    &  71.3(7) & 157.1(1.3) & 197.2(2)  &  120.5(3.1) \\
$\delta \nu_\mathrm{BBR}(^1\!S_0)$    & -0.614(6)& -1.353(11) &-1.698(2)  & -1.04(3)    \\
\hline
$\alpha_{^3\!P_0}$                    & 101.2(3) & 290.3(1.5) &458.3(3.6) &  266(15)    \\
$\delta \nu_\mathrm{BBR}(^3\!P_0)$    &-0.872(3) & -2.524(13) & -4.052(32)& -2.29(13)
\end{tabular}
\end{ruledtabular}
\end{table}

The results of calculations for the static electric dipole polarizabilties
for the $ns^2 \,^1\!S_0$ and $nsnp\, ^3\!P_0$ states are presented in Table~%
\ref{Tab:polariz}. The listed values of the ground-state
polarizabilities of Mg, Ca, and Sr were obtained by us
earlier~\cite{PorDer06Z}. To estimate their uncertainties we used
the fact that the intermediate state $nsnp\, ^1\!P_1$ contributes to
the polarizability at the level of 95-97\%. For calculating the
polarizabilities we used the best known in the literature values of
the $\langle ns^2 \,^1\!S_0 ||D|| nsnp\, ^1\!P_1 \rangle$ matrix
elements tabulated in~\cite{PorDer06Z}. For instance, for Sr
$|\langle 5s^2 \,^1\!S_0 ||D|| 5s5p\, ^1\!P_1 \rangle| = 5.249(2)$
a.u. leading to 0.1\% error in $\alpha_{^1\!S_0}$. The uncertainties
in the remaining polarizabilities were estimated as a half of the
difference between two predictions obtained with $\delta=0$ and with
$\delta$ determined with the best fit to the experimental energies
(thus mimicking omitted higher-order many-body corrections). The
uncertainties in the ground-state polarizabilities range from 0.1\%
for Sr to 3\% for Yb. For the $^3\!P_0$ states the errors range from
0.3\% for Mg to 6\% for Yb.

With the computed polarizabilities we can find the BBR frequency shifts with
Eq.(\ref{Eq:delEg}). The ``dynamic'' correction $\eta$ is negligible for the
$^1\!S_0$ states, but is needed for the $^3\!P_0$ calculations. Indeed, for
the ground state, the smallest excitation energy $%
E_{^{1}P_{1}^{o}}-E_{^{1}S_{0}}$ is equal to $21698$ cm$^{-1}$ for Sr. At $%
T=300\,\mathrm{K}$ the characteristic value of $y \sim 100$ for all the
atoms. By contrast, for the $^3\!P_0$ clock level, the transitions to the
nearby $^3\!D_1$ level involve smaller energies. For Sr, the relevant energy
is only $3841\, \mathrm{cm^{-1}}$ corresponding to characteristic value of $%
y\sim 20$. At this value, the ``static polarizability''
approximation, $F_1 (y) \approx 4\pi ^{3}/(45y)$, has only a few
percent accuracy. While evaluating $\eta$ we find it sufficient to
truncate the summation over intermediate states at the lowest-energy
excitation. This ``dynamic'' correction contributes to the BBR shift
of the $^3\!P_0$ state at 0.1\% level in Mg, 1\% in Ca, 2.7\% in Sr,
and 0.7\% in Yb. Notice that since the clock BBR shift is obtained
by subtracting BBR shifts of the individual levels, the ``dynamic''
correction contributes at an enhanced 5\% level in Sr. These
``dynamic'' corrections must be taken into account if the BBR shifts
are derived from dc Stark shift measurements.

Finally, we combine the BBR shifts of the individual clock levels and arrive
at the overall BBR corrections summarized in Table~\ref{Tab:BBRsum}. Our
computed BBR shift for Sr, $-2.354(32)$ Hz is in agreement
with an estimate~\cite{TakHinHig05} of $-2.4(1)$ Hz. Our uncertainties are
better than 3\%, except for Yb where the error is 10\%. As discussed in the
introduction, although resulting from state-of-the-art relativistic
atomic-structure calculations, these errors are still large and
substantially affect the projected $10^{-18}$ fractional accuracy of the
lattice-based clocks (see Table~\ref{Tab:BBRsum}). Potential solution
involves operating the clocks at cryogenic temperatures.

We would like to thank C. Oates and E. N. Fortson for motivating
discussions. This work was supported in part by the NSF Grant No.
PHY-0354876, NIST precision measurement grant, and RFBR Grant Nos.
04-02-16345-a and 05-02-16914-a.

\bibliographystyle{apsrev}


\end{document}